%% file: main.tex
\documentclass{article}
\usepackage{geometry}
\usepackage[utf8]{inputenc} 
\usepackage[T1]{fontenc}
\usepackage{url}
\usepackage{ifthen}
\usepackage{cite}
\usepackage{amsmath}
\usepackage{amsfonts}
\usepackage{amsthm}
\usepackage{mathtools}
\usepackage{amssymb}
\usepackage{xcolor}
\usepackage{tcolorbox}

\DeclarePairedDelimiterX{\bracket}[3]{#1}{#2}{#3}

\newcommand{\round}[1]{\bracket*{(}{)}{#1}}
\newcommand{\curly}[1]{\bracket*{\lbrace}{\rbrace}{#1}}
\newcommand{\squarebrack}[1]{\bracket*{\lbrack}{\rbrack}{#1}}
\newcommand{\intset}[1]{\squarebrack{#1}}
\newcommand{\set}[1]{\curly{#1}}
\newcommand{\prob}[1]{\mathbb{P}\round{#1}}
\newcommand{\expect}[1]{\mathbb{E}\squarebrack{#1}}

\newtheorem{theorem}{Theorem}
\newtheorem{definition}{Definition}
\newtheorem{lemma}{Lemma}

\newtheorem{conjecture}{Conjecture}

\usepackage{bbm}

\newcommand\blfootnote[1]{%
  \begingroup
  \renewcommand\thefootnote{}\footnote{#1}%
  \addtocounter{footnote}{-1}%
  \endgroup
}

\begin{document}
\title{Semidefinite Programming for the Asymmetric Stochastic Block Model}
\author{Julia Gaudio\footnote{Northwestern University Department of Industrial Engineering and Management Sciences}~ and Phawin Prongpaophan\footnote{Northwestern University Department of Computer Science}}

\maketitle

\begin{abstract}
   \input{abstract.tex}
   \blfootnote{J.G. and P.P. were supported in part by NSF CCF 2154100. P.P. was supported in part by NSF HDR TRIPODS (IDEAL Institute) and NSF CAREER Award CCF 2440539.}
\end{abstract}

\section{Introduction}

\input{introduction.tex}

\paragraph*{Notation} \input{notation.tex}

\paragraph*{Organization} \input{organization.tex}
 
\section{Model and Main Results} \label{section:model}

\input{model.tex}

\section{Proof of Theorem \ref{thm:main-result}} \label{section:proof}

\input{proof.tex}

\section{Discussion} \label{section:discussion}

\input{discussion.tex}

\pagebreak

\bibliographystyle{IEEEtran}
\bibliography{references}

\end{document}

%% file: abstract.tex
We consider semidefinite programming (SDP) for the binary stochastic block model with equal-sized communities. Prior work of Hajek, Wu, and Xu proposed an SDP (sym-SDP) for the symmetric case where the intra-community edge probabilities are equal, and showed that the SDP achieves the information-theoretic threshold for exact recovery under the symmetry assumption. A key open question is whether SDPs can be used to achieve exact recovery for non-symmetric block models. In order to inform the design of a new SDP for the non-symmetric setting, we investigate the failure of sym-SDP when it is applied to non-symmetric settings. We formally show that sym-SDP fails to return the correct labeling of the vertices in some information-theoretically feasible, asymmetric cases. In addition, we give an intuitive geometric interpretation of the failure of sym-SDP in asymmetric settings, which in turn suggests an SDP formulation to handle the asymmetric setting. Still, this new SDP cannot be readily analyzed by existing techniques, suggesting a fundamental limitation in the design of SDPs for community detection.

%% file: introduction.tex

The Stochastic Block Model (SBM) \cite{Holland1983} is considered the canonical probabilistic generative model for graphs with community structure. Since its introduction in 1983, the SBM has attracted significant attention from the probability, statistics, machine learning, and information theorety communities. The SBM with $k$ communities is specified by a community membership vector $\pi \in [0,1]^k$ satisfying $\sum_{i \in [k]} \pi_i = 1$ and a symmetric edge probability matrix $P \in [0,1]^{k \times k}$. A graph $G \sim \text{SBM}(n, \pi, P)$ is sampled by first independently assigning each vertex $v \in [n]$ to a community $\sigma^{\star}(v)$, where $\mathbb{P}(\sigma^{\star}(v) = i) = \pi_i$ for all $i \in [k]$. Conditioned on the community assignments, edges are sampled independently, with $\mathbb{P}(\{u,v\} \in E) = P_{\sigma^{\star}(u), \sigma^{\star}(v)}$. Given the unlabeled graph $G$, the goal is to recover the community labels $\sigma^{\star}(\cdot)$. Community detection in the SBM has been studied in both dense and sparse regimes, with different inference goals: exact, almost exact, and partial recovery. For an overview of recent developments, see the survey of Abbe \cite{Abbe2017}.

A variety of algorithms have been proposed for community detection, including spectral algorithms \cite{Mcsherry2001,Vu2018,Yun2014,Lelarge2015,Chin2015,Krzakala2013,Massoulie2014,Bordenave2015,Abbe2020}, graph splitting \cite{Abbe2015}, belief propagation \cite{Decelle2011,Abbe2015b}, and semidefinite programming (SDP), which is the focus of our paper. In particular, we are interested in the exact recovery problem, for which the information-theoretic threshold was determined by Abbe and Sandon \cite{Abbe2015}.

There is a substantial body of work on SDPs for community detection. Semidefinite programming was first proposed for community recovery in the two community setting by Feige and Kilian \cite{Feige2001}, though their algorithm did not achieve the information-theoretic threshold for exact recovery. Later, Abbe, Bandeira, and Hall \cite{Abbe2015c} proposed an SDP for exact recovery in the balanced, symmetric SBM with two communities (with $\pi_1 = \pi_2 = 1/2$ and $P_{11} = P_{22}$), and conjectured it to be optimal. This conjecture was affirmatively resolved by Hajek, Wu, and Xu \cite{Hajek2016} with a slightly modified SDP. The same authors additionally showed that SDPs are optimal for exact recovery in several other inference problems: the symmetric but unbalanced SBM, the SBM with $k \geq 3$ communities of equal size with all inter- and intra-community probabilities equal, and the Censored SBM (in which only some edge observations are available) \cite{Hajek2016b}, the Planted Dense Subgraph Model (where $P_{12} = P_{21} = P_{22}$) and Submatrix Localization \cite{Hajek2016c}. Perry and Wein \cite{Perry2017} subsequently developed an SDP for the case of multiple unbalanced communities with all inter- and intra-community probabilities equal. In light of these developments, an important question stands open:
\begin{center}
    \emph{Do SDPs achieve exact recovery in general SBMs?}
\end{center}

In this paper, we provide evidence to highlight the challenge in designing an optimal SDP algorithm by showing that the SDP introduced in \cite{Hajek2016} fails on some asymmetric cases. Our analysis uncovers a geometric interpretation of the failure. Using this geometric insight, we propose a new SDP for asymmetric cases. Intriguingly, the new SDP is not amenable to standard analysis tools, which suggests that in fact SDPs do not achieve the IT threshold for exact recovery.

\subsection{Further related work}
SDPs have additionally been proposed for weak, partial, and almost exact recovery \cite{Fei19, Montanari16, pmlr-v99-fei19a}. Gu{\'e}don and Vershynin\cite{Guedon2016} introduced a technique based on Grothendieck's inequality, which provides guarantees for SDPs to achieve almost exact recovery.

While SDPs are much slower than other approaches such as spectral clustering, they are known to be more robust \cite{Feige2001,Hajek2016,CL15,pmlr-v49-makarychev16}. For example, for the problem of exact recovery in the symmetric, balanced SBM, the sym-SDP formulation is robust to monotone adversaries (which can add intra-community edges and remove inter-community edges) \cite{Feige2001,Hajek2016}. While the monotonicity appears to make the recovery problem easier, other algorithms are brittle to such changes. Interestingly, while a monotone adversary does not change the IT threshold for exact recovery in the symmetric, balanced SBM, such adversaries \emph{do} change the IT threshold for partial recovery in the sparse SBM \cite{Moitra2016}.

There has been limited work on showing impossibility of recovery using SDPs, above the IT threshold. A notable work in this direction is the paper of Amini and Levina \cite{Amini2018}, who showed that strong assortativity (which requires $\min\round{P_{11}, P_{22}} > P_{12}$) is a necessary condition for the SDP formulations of \cite{CXJ16}, \cite{CL15} to achieve exact recovery down to the IT threshold.
Related work of Gaudio, Dhara, Mossel, and Sandon \cite{Dhara2023} shows that spectral algorithms using a single matrix encoding fail to achieve the IT threshold for exact recovery in the censored SBM (in which edge observations are missing at random), and provides a geometric interpretation of the failure in terms of the geometry of \emph{degree profiles}, which we draw on in this paper.

Apart from community detection, SDPs have been used for other combinatorial problems such as Max-Cut \cite{GW94}, coloring \cite{KMS94}, ranking \cite{Yang2024TopKRW}, and clustering \cite{pmlr-v99-fei19a}.

%% file: notation.tex
We define $A$ to be the adjacency matrix of the input graph $G$.
For any vector $v$, let $v(i)$ denote the $i$-th entry of $v$.
For any positive integer $n$, let $\squarebrack{n} = \set{1, 2, \ldots, n}$.
For any set $S$, let $|S|$ denotes its cardinality and let $S^C$ denote its complement.
We use $\mathbf{I}$ for the identity matrix, $\mathbf{J}$ for the all-ones matrix and $\mathbf{1}$ for the all-ones vector, with the appropriate dimensions.
For any symmetric matrices $A$ and $B$ of the same size, let $\langle A, B \rangle = \mathrm{tr}\round{AB}$ be their inner product.
We use $\textsf{Poi}\round{\lambda ; k}$ to denote the probability that a Poisson random variable with parameter $\lambda$ takes value $k$.
All logarithms are natural.
We use standard asymptotic notation.

%% file: organization.tex
The rest of the paper is organized as follows.
We introduce the SBM model and Hajek, Wu, and Xu's SDP along with our main result in Section \ref{section:model}, while the proof of our result is presented in Section \ref{section:proof}. Finally, we conclude with discussion in Section \ref{section:discussion}.

%% file: model.tex

Let $n$ be an even number, and let $V = \intset{n}$ be a set of vertices which is partitioned into two equal-sized communities $C_1$ and $C_2$.
Let $\sigma^\star \in \set{\pm 1}^n$ denote the community membership vector, where

\begin{equation*}
    \sigma^\star \round{u} = \begin{cases}
        1 & u \in C_1 \\
        -1 & u \in C_2.
    \end{cases}
\end{equation*}

A random graph $G \sim \textsf{SBM}\round{n, \alpha_1, \alpha_2, \beta}$ is generated by independently sampling edges. For each pair of vertices $u \neq v$, edges are generated with the following probability, conditioned on $\sigma^{\star}$: 
\begin{equation*}
    \prob{\round{u, v} \in E} = \begin{cases}
        p_1 = \alpha_1 \cdot \frac{\log n}{n} & \sigma^\star\round{u} = \sigma\round{v} = 1 \\
        p_2 = \alpha_2 \cdot \frac{\log n}{n} & \sigma^\star\round{u} = \sigma\round{v} = -1 \\
        q = \beta \cdot \frac{\log n}{n} & \sigma^\star\round{u} \neq \sigma\round{v}.
    \end{cases}
\end{equation*}
Here, $p_i$ is the probability of intra-community edges for cluster $C_i$ while $q$ is the probability of inter-community edges.

Abbe and Sandon \cite{Abbe2015} determined the information-theoretic threshold for exact recovery. In the special case of two communities of equal size, the threshold is stated in terms of the function $\textsf{IT}\round{\alpha_1, \alpha_2, \beta}$, which equals
\begin{equation} \label{eq:it-threshold}
\textsf{IT}\round{\alpha_1, \alpha_2, \beta} = \sup_{t \in \squarebrack{0, 1}} \frac{1}{2} \squarebrack{t \alpha_1 + \round{1 - t} \alpha_2 + \beta  - \beta^t \alpha_2^{(1 - t)} - \beta^{(1 - t)} \alpha_1^t}.
\end{equation}
If $\textsf{IT}\round{\alpha_1, \alpha_2, \beta} > 1$ then the communities can be exactly recovered (up to permutation of labeling) with probability converging to one as $n \rightarrow \infty$. On the other hand, if $\textsf{IT}\round{\alpha_1, \alpha_2, \beta} < 1$ then no algorithm can exactly recover the communities with probability converging to one.

The threshold (\ref{eq:it-threshold}) is achieved by the maximum likelihood estimator (MLE), which finds the labeling $\sigma$ with the highest likelihood. 
In the special case of $p_1 = p_2 > q$, the MLE of $\sigma^\star$ can be simply stated as
\begin{equation}
\begin{array}{ll@{}ll}
\hat{\sigma}_{\mathrm{MLE}} = \arg\max_{\sigma}  &  z(\sigma) \equiv \sum_{i, j} A_{ij} \sigma(i) \sigma(j) &\\
\text{subject to}& \sigma \in \set{\pm 1}^n, \sigma^\top \mathbf{1} = 0
\end{array}
\end{equation}
The optimization problem above is known to be NP-hard. Hajek, Wu, and Xu \cite{Hajek2016} introduced the following convex relaxation:
\begin{equation} \label{eq:hwx-sdp}
\begin{array}{ll@{}ll}
 \hat{Y}_{\mathrm{SDP}} = \arg\max_Y  &  \langle A, Y \rangle &\\
\text{subject to}& Y \succeq 0 \\
& Y_{ii} = 1, \forall i \in \intset{n} \\
& \langle \mathbf{J}, Y \rangle = 0,
\end{array}
\end{equation}
With high probability, $X^\star = {\sigma^\star} \sigma^{\star^\top}$ is the \emph{unique} solution to the SDP (\ref{eq:hwx-sdp}) under the input graph $G \sim \textsf{SBM}\round{n, \alpha_1, \alpha_2, \beta}$ where $\alpha_1 = \alpha_2$; thus, the SDP recovers $\sigma^{\star}$ in this special case \cite{Hajek2016}.

Hajek, Wu, and Xu's result provides a fundamental building block for designing SDPs that mimic the MLE. Hence, the first step toward designing SDPs for other cases would be understanding the limitation of the existing approach for parameter values that it was not designed for.

Our result formally shows that the SDP (\ref{eq:hwx-sdp}) designed for a special case of $p_1 = p_2 > q$ fails to return an optimal solution in a general case of $p_1 \neq p_2$. 

\begin{theorem}\label{thm:main-result}
    There exist parameters $\round{\alpha_1, \alpha_2, \beta}$ such that $\textsf{IT}\round{\alpha_1, \alpha_2, \beta} > 1$ but $X^\star = {\sigma^\star} \sigma^{\star^\top}$ is not the optimal solution to the SDP in (\ref{eq:hwx-sdp}) with probability at least $1 - o(1)$.
\end{theorem}

Note that Theorem \ref{thm:main-result} only states that the target solution $X^\star = {\sigma^\star} \sigma^{\star^\top}$ is suboptimal with high probability. In particular, this result does not rule out the possibility of obtaining the correct labeling by rounding the decomposition of $\hat{Y}_{\mathrm{SDP}}$.
Although it is not surprising that SDP (\ref{eq:hwx-sdp}) fails on the input they are not designed for, our analysis provides geometric intuition on its failure and gives suggestions on how to modify the SDP.

%% file: proof.tex

\subsection{Proof Overview}

To show that $X^{\star} = \sigma^\star \sigma^{\star \top}$ is suboptimal, we will show that there exists another solution $X = \sigma \sigma^{\top}$ with $\sigma \neq \sigma^\star$ that gives a higher objective value. We will limit our search to such $\sigma$ that can be obtained by swapping the label of one pair of vertices relative to the ground truth. More formally, for $i \in C_1$ and $j \in C_2$, let $\sigma^\star_{i \leftrightarrow j}$ be the labeling where
\begin{equation*}
    \sigma^\star_{i \leftrightarrow j}(u) = \begin{cases}
        -1 & u = i \\
        1 & u = j \\
        \sigma^\star(u) & u \in \intset{n} \setminus \set{i, j}.
    \end{cases}
\end{equation*}
For $i \in C_1$ and $j \in C_2$, define $E_{ij}$ to be the event where $z\round{\sigma^\star_{i \leftrightarrow j}} > z\round{\sigma^\star}$. That is, the objective value increases if we swap the labels of vertices $i$ and $j$ from the ground truth.
\begin{align}
     E_{ij} &= \set{z\round{\sigma^\star_{i \leftrightarrow j}} > z(\sigma^\star)}
    = \set{z\round{\sigma^\star_{i \leftrightarrow j}} - z(\sigma^\star) > 0} \nonumber \\
    &= \set{\sum_{\ell \in C_2} A_{i\ell} + \sum_{\ell \in C_1} A_{j \ell} - \sum_{\ell \in C_1} A_{i \ell} - \sum_{\ell \in C_2} A_{j \ell} > 0} \nonumber \\
    &= \set{\round{\sum_{\ell \in C_2} A_{i \ell} - \sum_{\ell \in C_1} A_{i \ell}} + \round{\sum_{\ell \in C_1} A_{j \ell} - \sum_{\ell \in C_2} A_{j \ell}} > 0} \label{eq:failure-event}
\end{align}

Notice that the first term relates to vertex $i$ and the second relates to vertex $j$. We will define parameterized events $\set{F_i(\delta)}_i$ and $\set{G_j(\varepsilon)}_j$ as follows:
\begin{align}
    F_i(\delta) &= \set{\sum_{\ell \in C_1} A_{i\ell} - \sum_{\ell \in C_2} A_{i \ell} \leq \delta \log n}, \; i \in C_1 \label{eq:def-event-f} \\
    G_j(\varepsilon) &= \set{\sum_{\ell \in C_2} A_{j\ell} - \sum_{\ell \in C_1} A_{j \ell} \leq -\varepsilon \log n}, \; j \in C_2. \label{eq:def-event-g}
\end{align}

The events are designed in a way that when $\varepsilon > \delta$, $\round{F_i(\delta) \cap G_j(\varepsilon)} \subseteq E_{ij}$,  which implies that $\sigma^\star$ is not an optimal solution. Our goal is to show that $\cup_{i \in C_1, j \in C_2} E_{ij}$ happens with high probability.

To achieve that goal, we will show that $\prob{\cup_{i \in C_1} F_i\round{\delta}} = 1 - o(1)$ and $\prob{\cup_{j \in C_2} G_j\round{\varepsilon}} = 1 - o(1)$, for some $\varepsilon > \delta > 0$.
A difficulty arises due to the fact that $\{F_i \round{\delta}\}_{i \in C_1}$ and $\{G_j \round{\varepsilon}\}_{j \in C_2}$ are dependent. Using standard techniques (see \cite[Eq. 32]{Hajek2016}), they can essentially be treated as independent.
If these events were indeed independent, and $\prob{F_i \round{\delta}}, \prob{G_j \round{\varepsilon}} = \omega\round{1 / n}$ then $\prob{\cup_{i \in C_1} F_i \round{\delta}} = 1 - \prob{\cap_{i \in C_1} F_i \round{\delta} ^C} = 1 - \squarebrack{1 - \omega\round{1 / n}}^{n / 2} = 1 - o(1)$.
By the same reasoning, $\prob{\cup_{j \in C_2} G_j \round{\varepsilon}} = 1 - o(1)$. Hence, $\prob{\cup_{i \in C_1, j \in C_2} \round{F_i \round{\delta} \cap G_j \round{\varepsilon}}} = 1 - o(1)$. We will later address dependence  by introducing proxy events $\bar{F}_i \round{\delta}$ and $\bar{G}_j \round{\varepsilon}$ (Section \ref{section:main-proof}).

\vspace{1em}


The events $F_i \round{\delta}$ and $G_j \round{\varepsilon}$ are in the form of tail bounds on vertices $i$ and $j$. It turns out that these tail bounds can be characterized through the geometry of so-called \emph{degree profiles}, defined below:

\begin{definition}[Degree Profile]
    The (rescaled) degree profile of vertex $u \in \intset{n}$ is given by \[d_u = \round{\frac{\sum_{\ell \in C_1} A_{u\ell}}{\log n}, \frac{\sum_{\ell \in C_2} A_{u\ell}}{\log n}} \in \mathbb{R}^2.\]
\end{definition}

In terms of degree profiles, our event $F_i \round{\delta}$ is precisely the event that $w^{\star \top} d_i \leq \delta$ where $w^\star = \round{1, -1}^\top$. Also, $G_j \round{\varepsilon}$ is the event where $w^{\star \top} d_j \geq \varepsilon$.
The essence of the inner product $w^{\star \top} d_u$ comes from its relationship to the so-called \emph{Genie-Aided Estimator}. To make the connection precise, consider the related SBM model where the label of each vertex is independent and uniform on $\{\pm 1\}$, and suppose that a genie reveals all labels except for that of vertex $u$. By deriving the MAP estimator for $u$ given the genie information, the optimal estimator of $\sigma^{\star}(u)$ is given by $\hat{\sigma}\round{u} = \mathrm{sign}\squarebrack{w^{\star ^\top} d_u}$.
Thus, the genie's success requires $\min_{i \in C_1} w^{\star \top} d_i > \max_{j \in C_2} w^{\star \top} d_j$, a condition that fails if $F_i(\delta) \cap G_j(\varepsilon)$ holds for $\varepsilon > \delta$ and some $i \in C_1, j \in C_2$.
Moreover, the genie condition is also tied to the success of the dual certificate proof for showing the optimality of the sym-SDP in the symmetric setting, which amounts to showing that $\sum_{\ell = 1}^n A_{i\ell} \sigma^\star_i \sigma^\star_\ell \geq \eta \log n$ for all $i \in [n]$ and some $\eta > 0$ \cite{Hajek2016}.

\begin{figure}[h] 
\centering
\vspace{-1em}
\includegraphics[width=0.6\textwidth]{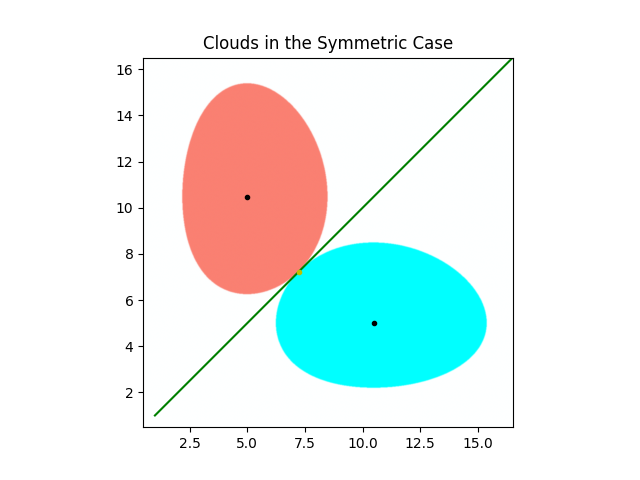}
\caption{Visualization of the clouds corresponding to $C_1$ (blue) and $C_2$ (red) for $\beta = 10, \alpha_1 = \alpha_2 \approx 20.94$. The separating line $w^{\star \top} x = 0$ is marked in green. This is exactly the case where $\textsf{IT}\round{\alpha_1, \alpha_2, \beta} = 1$ where the two clouds touch at one single point (shown in yellow).}
\label{fig:cloud}
\end{figure}

\begin{figure}[h] 
\centering
\vspace{-1em}
\includegraphics[width=0.6\textwidth]{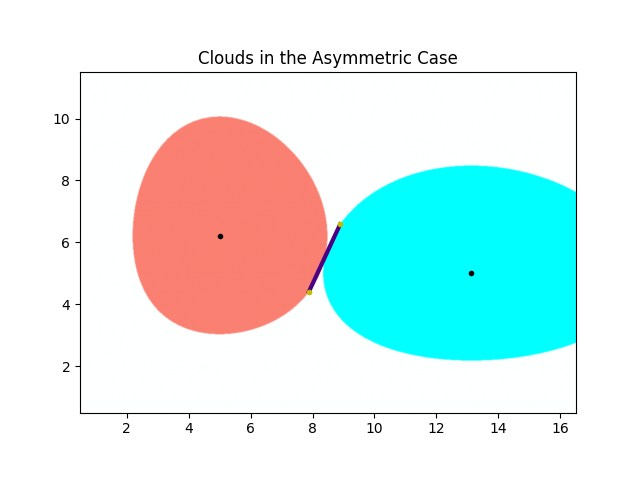} 
\caption{Visualization of the clouds corresponding to $C_1$ (blue) and $C_2$ (red) for $\beta = 10, \alpha_1 \approx 26.24, \alpha_2 \approx 12.43$ so that $\textsf{IT}\round{\alpha_1, \alpha_2, \beta} = 1$.
The purple line has slope $2.16 > 1$ and connects two degree profiles from different clouds so that $w^\star$ cannot separate them. Note that when the parameters change slightly to make $\textsf{IT}\round{\alpha_1, \alpha_2, \beta} > 1$, the slope will be slightly changed, and the argument about the failure of separation still holds.}
\label{fig:cloud-slope}
\end{figure}

The set of degree profiles which occur with probability $\omega(1/n)$ for a given vertex in a particular community forms a \emph{cloud} as shown in Figure \ref{fig:cloud}.
The IT feasibility ensures that two clouds, corresponding to the two communities, are not overlapping so that the Genie-aided estimator can distinguish vertices from different communities. In particular, the case where $\textsf{IT}\round{\alpha_1, \alpha_2, \beta} = 1$ is exactly the case where the two clouds touch at one single point.
The success of the Genie-Aided Estimator in the symmetric case comes from the fact that those two clouds are separable by $w^\star$ (visualized with a green line in Figure \ref{fig:cloud}). When the SBM parameters change, $w^\star$ may be no longer be the optimal separator. Formally, suppose that there are two degree profiles $\round{x_1, y_1}$ in Community 1, and $\round{x_2, y_2}$ in Community 2 such that the slope of the line connecting them is strictly larger than $1$; i.e., $\frac{y_1 - y_2}{x_1 - x_2} > 1$. In that case, $w^{\star \top} (x_1, y_1) < w^{\star \top} (x_2, y_2)$, and so $w^{\star}$ fails to separate the degree profiles (see Figure \ref{fig:cloud-slope}).

In fact, the slope condition of degree profile is asymptotically equivalent to the existence of $\varepsilon > \delta$ required for $\round{F_i(\delta) \cap G_j(\varepsilon)} \subseteq E_{ij}$ 
To conclude the failure of the SDP, we will show that there exist IT-feasible parameters that admit the witness points that satisfy the slope condition, and at the same time are likely to arise.


\subsection{Main Proof} \label{section:main-proof}
As identified earlier, the events $F_i \round{\delta}$ and $G_j \round{\varepsilon}$ defined in (\ref{eq:def-event-f}) and (\ref{eq:def-event-g}) are dependent. To resolve the issue, we will construct independent, closely related events. For $i \in \set{1, 2}$, fix $T_{i} \subseteq C_i$ such that $|T_i| = \frac{n}{\log^3 n}$. For example, $T_i$ can be the set of $\frac{n}{\log^3 n}$ vertices in $C_i$ with smallest index.  Also define $C_i' = C_i \setminus T_i$.

Then, we define a likely event $L$ as follows: 
\begin{equation*}
     L = \set{\max_{i \in T_1} \round{\sum_{\ell \in T_1} A_{i \ell}} \leq 1} \cap \set{\max_{j \in T_2} \round{\sum_{\ell \in T_2} A_{j \ell}} \leq 1}
\end{equation*}

\begin{lemma} \label{lemma:likely-prob}
    $\prob{L} = 1 - o(1/n)$.
\end{lemma}

We introduce new parameterized events $\bar{F}_i \round{\delta}$ and $\bar{G}_j \round{\varepsilon}$ as follows:
\begin{align*}
    \bar{F}_i \round{\delta} &= \set{\sum_{\ell \in C_1'} A_{i\ell} - \sum_{\ell \in C_2'} A_{i \ell} + 1 < \delta \log n},\; i \in T_1  \\
    \bar{G}_j \round{\varepsilon} &= \set{\sum_{\ell \in C_2'} A_{j\ell} - \sum_{\ell \in C_1'} A_{j \ell} + 1 < -\varepsilon \log n},\; j \in T_2.
\end{align*}

Notice that they differ from the previous $F_i \round{\delta}$ and $G_j \round{\varepsilon}$ by the set of indices in the summation and the additional buffer of $1$ that accounts for missing indices in the summation. But now, $\bar{F}_i \round{\delta}$ and $\bar{G}_j \round{\varepsilon}$ are independent conditioned on the community assignments.

By the construction, we can see that $\round{L \cap \bar{F}_i \round{\delta}} \subseteq F_i \round{\delta}$ and $\round{L \cap \bar{G}_j \round{\varepsilon}} \subseteq G_j \round{\varepsilon}$.
A standard analysis argument implies an analogous relation in the other direction, as stated in the following lemma, whose proof is deferred.

\begin{lemma}\label{lemma:newEventToBasicSum}
For all $\delta' > \delta$, $\prob{F_i \round{\delta} \mid L} \leq \prob{\bar{F}_i \round{\delta'} \mid L}$ for sufficiently large $n$. Similarly, for all $\varepsilon' < \varepsilon$, $\prob{G_j \round{\varepsilon} \mid L} \leq \prob{\bar{G}_j \round{\varepsilon'} \mid L}$ for sufficiently large $n$.
\end{lemma}

Both directions of the relation suggest that we can use the probabilities of the original events $F_i$ and $G_j$ as lower bounds of probabilities of the new events $\bar{F_i}$ and $\bar{G}_j$ (with appropriate parameters). As mentioned earlier, these events $F_i \round{\delta'}$ and $G_j \round{\varepsilon'}$ are closely related to the degree profile and for any degree profile $d_u$, we can use the value of $w^{\star \top} d_u$ to check whether it belongs to events $F_i \round{\delta'}$ and/or $G_j \round{\varepsilon'}$.

The formal relations between the chosen parameters and the failure of the SDP will be described in the following lemma.

\begin{lemma} \label{lemma:witness-to-failure}
    If there exists parameters $0 < \delta < \varepsilon$ such that $\prob{F_i \round{\delta}} \geq n^{-c}$ and $\prob{G_j \round{\varepsilon}} \geq n^{-c}$ for some $c < 1$, then $X^\star = {\sigma^\star} \sigma^{\star \top}$ is not the optimal solution to the SDP in (\ref{eq:hwx-sdp}) with probability $1 - o(1)$.
\end{lemma}

Finally, we prove Theorem \ref{thm:main-result} by showing that the condition in Lemma \ref{lemma:witness-to-failure} is satisfied for some IT-feasible parameters.
\begin{proof}[Proof of Theorem \ref{thm:main-result}]
Notice that the degree profile $d_u$ is a vector of two binomial random variables, which can be well-approximated by Poisson random vectors (see e.g. \cite[Lemma 23]{Dhara2023}). We have that for $i \in C_1$ and $j \in C_2$,
\begin{align}
    \prob{d_i = \round{x, y}} &\asymp \textsf{Poi}\round{\frac{\alpha_1}{2} \log n ; x \log n} \; \textsf{Poi}\round{\frac{\beta}{2} \log n ; y \log n} \nonumber \\
    &= \frac{\round{\frac{\alpha_1}{2} \log n}^{x \log n} \exp\round{-\frac{\alpha_1}{2} \log n} }{\round{x \log n}!}
    \cdot
    \frac{\round{\frac{\beta}{2} \log n}^{y \log n} \exp\round{-\frac{\beta}{2} \log n} }{\round{y \log n}!} \nonumber
    \intertext{Using Stirling's approximation,}
    &\sim \frac{\round{\frac{e \cdot \frac{\alpha_1}{2} \log n}{x \log n}}^{x \log n} \exp\round{-\frac{\alpha_1}{2} \log n}}{\sqrt{2 \pi x \log n}} \cdot 
    \frac{\round{\frac{e \cdot \frac{\beta}{2} \log n}{y \log n}}^{y \log n} \exp\round{-\frac{\beta}{2} \log n}}{\sqrt{2 \pi y \log n}} \nonumber \\
    &= \frac{1}{2 \pi \log n \sqrt{xy}} \cdot
    \round{\frac{\alpha_1 e}{2 x}}^{x \log n}
    \exp\round{- \frac{\alpha_1}{2} \log n}
    \round{\frac{\beta e}{2 y}}^{y \log n}
    \exp\round{- \frac{\beta}{2} \log n} \nonumber \\
    &\asymp \mathrm{pow}\squarebrack{n, o(1) - \frac{1}{2} \round{\alpha_1 + \beta} + x \log \round{\frac{\alpha_1 e}{2 x}} + y \log \round{\frac{\beta e}{2 y}} } \label{eq:pow-form-1}
    \intertext{These steps can also be applied to $d_j$ so that}
    \prob{d_j = \round{x, y}} &\asymp \textsf{Poi}\round{\frac{\beta}{2} \log n ; x \log n} \; \textsf{Poi}\round{\frac{\alpha_2}{2} \log n ; y \log n} \nonumber \\
    &= \mathrm{pow}\squarebrack{n, o(1) - \frac{1}{2} \round{ \beta + \alpha_2} + x \log \round{\frac{\beta e}{2 x}} + y \log \round{\frac{\alpha_2 e}{2 y}} }. \label{eq:pow-form-2}
\end{align}
The exponents of expressions (\ref{eq:pow-form-1}) and (\ref{eq:pow-form-2}) correspond to the value of $-c$ in Lemma \ref{lemma:witness-to-failure}.


In fact, the condition in Lemma \ref{lemma:witness-to-failure} has a geometric interpretation in terms of the slope condition of degree profiles. Namely, define two \emph{clouds} $\mathcal{C}_i = \set{d \in \mathbb{R}^2 \mid \prob{d_u = d} = \omega\round{1 / n}}$ for $u \in C_i$. Suppose there exists $d_1 = \round{x_1, y_1} \in \mathcal{C}_1$ and $d_2 = \round{x_2, y_2} \in \mathcal{C}_2$ such that $\frac{y_1 - y_2}{x_1 - x_2} > 1$, the slope condition implies that $x_2 - y_2 > x_1 - y_1$. Since the inequality is strict, there must be $\varepsilon > \delta$ such that $x_2 - y_2 > \varepsilon > \delta >  x_1 - y_1$.
That is, $w^{\star \top} d_1 < \delta$ and $w^{\star \top} d_2 < -\varepsilon$. With the probability requirement in the definition of $\mathcal{C}_1$ and $\mathcal{C}_2$, this is exactly the condition in Lemma \ref{lemma:witness-to-failure}.

Notice that the slope condition here is equivalent to the condition of impossibility of using $w^\star$ as a separating plane described earlier. Therefore, Figure \ref{fig:cloud-slope} which shows witness points in both clouds that are connected by a line with a slope strictly greater than one is already sufficient to conclude the proof of Theorem \ref{thm:main-result} by showing that the conditions in Lemma \ref{lemma:witness-to-failure} can be satisfied.
\end{proof}

\begin{figure}[h] 
\centering
\includegraphics[width=0.60\textwidth]{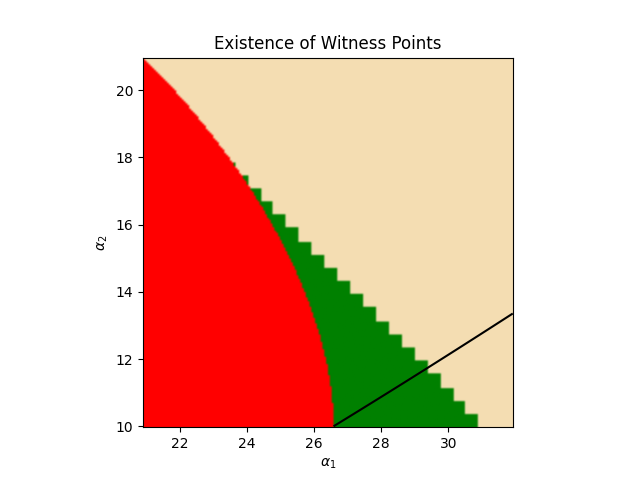} 
\caption{Parameters with possible witness points when $\beta = 10$. The red region is the IT-infeasible regime, and the green region is the set of parameters with possible witness points.
The parameters below the black line are achievable through the SDP for the PDS, using monotonicity.}
\label{fig:witness-region}
\end{figure}

In fact, the parameters chosen for Figure \ref{fig:cloud-slope} are not particularly special as there are many IT-feasible parameters that admit witness points that satisfy the conditions. For a fixed $\beta = 10$, a subset of choices of $\round{\alpha_1, \alpha_2}$ which satisfy the requirements are shown in Figure \ref{fig:witness-region}. Also note that some (but not all) of these parameters with witness points are achievable through the SDP for the Planted Dense Subgraph (PDS) Model using a monotonicity argument. Specifically, consider the SBM with parameters  $(\alpha_1, \alpha_2, \alpha_2)$, which is a special case of the PDS where the dense community is comprised of $n/2$ vertices. If $\alpha_1, \alpha_2$ are feasible parameters, then
the SDP derived for the PDS \cite[Equation 9]{Hajek2016} will achieve exact recovery. Let $A$ be sampled from the SBM with parameters $(\alpha_1, \alpha_2, \alpha_2)$ and let $A'$ be sampled from the SBM with parameters $(\alpha_1, \alpha_2, \beta)$, where $\beta < \alpha_2$. We can couple $A$ and $A'$ to ensure that $A \geq A'$ entrywise. By \cite[Lemma 4]{Hajek2016}, the maximizer of the SDP with objective $\langle A, Y \rangle$ is unique. Letting $Y^{\star}$ denote the maximizer, we have that for all feasible $Y \neq Y^\star$,
\begin{equation*}
    \langle A', Y \rangle
    = \langle A, Y \rangle - \langle A - A', Y \rangle
    \stackrel{(a)}{\leq} \langle A, Y \rangle
    \stackrel{(b)}{<} \langle A, Y^\star \rangle
    = \langle A', Y^\star \rangle.
\end{equation*}
Here, (a) holds as both $\round{A - A'}$ and $Y$ are non-negative and (b) holds by uniqueness of the optimizer $Y^\star$ under input $A$. This shows that $Y^\star$ is still a unique solution of the SDP designed for the PDS case when some intercommunity edges are removed.

\subsection{Proofs of Helper Lemmas}
\begin{proof}[Proof of Lemma \ref{lemma:likely-prob}]

Let $L_1 = \set{\max_{i \in T_1} \round{\sum_{\ell \in T_1} A_{i \ell}} \leq 1}$ be the first set of requirements of the event $L$. We will only show the analysis for $\prob{L_1}$ as the same reasoning also applies to the other half.

For $i \in T_1$, let $X_i = \sum_{\ell \in T_1} A_{i \ell}$ so that $X_i \sim \textsf{Bin}\round{\frac{n}{\log^3 n}, \alpha_1 \cdot \frac{\log n}{n}}$. Define \emph{bad} events $Q_i = \set{X_i \geq 1}$ so that $\prob{L_1} = 1 - \prob{\cup_{i \in T_1} Q_i}$.
For sufficiently large $n$ where $1 \geq 2 \expect{X_i}$, a Chernoff bound yields
\begin{align*}
    \prob{Q_i} = \prob{X_i \geq 1} &\leq \prob{X_i \geq 2 \expect{X_i}}
    \leq \exp \round{-\frac{1}{3} \expect{X_i}} \\
    &= \exp \round{-\frac{1}{3} \alpha_1 \log^{2} n}
    = n^{-\frac{1}{3} \alpha_1 \log n}.
\end{align*}
By a union bound, $\prob{L_1} \geq 1 - n^{-\frac{1}{3} \alpha_1 \frac{n}{\log^2 n}}$. This can be combined with $\prob{\max_{j \in T_2} \round{\sum_{\ell \in T_2} A_{j \ell}} \leq 1}$ using standard arguments to yield the desired result.
\end{proof}

\vspace{1em}

\begin{proof}[Proof of Lemma \ref{lemma:newEventToBasicSum}]
    We prove the statement about $F_i \round{\delta}$ by showing that $\round{F_i \round{\delta} \cap L} \subseteq \round{\bar{F}_i \round{\delta'} \cap L}$.
    On the event $\round{F_i \round{\delta} \cap L}$, we have
    \begin{align*}
        \sum_{\ell \in C_1} A_{i\ell} - \sum_{\ell \in C_2} A_{i \ell} &\leq \delta \log n \\
        \sum_{\ell \in C_1'} A_{i\ell} - \sum_{\ell \in C_2'} A_{i \ell}
        &\leq \delta \log n - \sum_{\ell \in T_1} A_{i\ell} + \sum_{\ell \in T_2} A_{i \ell} \\
        &\leq \delta \log n + 1 \leq \delta' \log n - 1,
    \end{align*}
    for $n$ sufficiently large.
    This set containment allows us to conclude that $\prob{F_i \round{\delta} \cap L } \leq \prob{\bar{F}_i \round{\delta'} \cap L}$, which implies $\prob{F_i \round{\delta} \mid L} \leq \prob{\bar{F}_i \round{\delta'} \mid L}$.
    By the same reasoning, we can conclude that $\prob{G_j \round{\varepsilon} \mid L} \leq \prob{\bar{G}_j \round{\varepsilon'} \mid L}$ as well.
\end{proof}

\vspace{1em}

\begin{proof}[Proof of Lemma \ref{lemma:witness-to-failure}]
    Let $0 < \delta < \varepsilon$ and $c < 1$ such that $\prob{F_i \round{\delta}} \geq n^{-c}$ and $\prob{G_j \round{\varepsilon}} \geq n^{-c}$. Since $\delta < \varepsilon$, there must exist $\delta'$ and $\varepsilon'$ such that $\delta < \delta' < \varepsilon' < \varepsilon$.

The failure probability is lower-bounded by
\begin{align}
     \prob{\bigcup_{i \in C_1, j \in C_2} E_{ij}} &\geq \prob{\round{\bigcup_{i \in C_1} F_i \round{\delta'}} \cap \round{\bigcup_{j \in C_2} G_j \round{\varepsilon'}} } \nonumber \\
    &\geq \prob{\round{\bigcup_{i \in T_1} \bar{F}_i \round{\delta'}} \cap \round{\bigcup_{j \in T_2} \bar{G}_j \round{\varepsilon'}} \cap L} \nonumber \\
    &= \prob{ \bigcup_{i \in T_1} \bar{F}_i \round{\delta'} \mid L} \prob{ \bigcup_{j \in T_2} \bar{G}_j \round{\delta'} \mid L} \prob{L} \label{eq:independent-breaker}
\end{align}
Let $u \in C_1$. Let $c < c' < 1$, and consider the first factor:
\begin{align*}
    \prob{ \bigcup_{i \in T_1} \bar{F}_i \round{\delta'} \mid L} &= 1 - \prob{\bigcap_{i \in T_1} \bar{F}_i^C \round{\delta'} \mid L} \\
    &= 1 - \prob{\bar{F}_u^C \round{\delta'} \mid L }^{\frac{n}{\log^3 n} } \\
    &\overset{(a)}{\geq}  1 - \prob{F_u^C \round{\delta} \mid L }^{\frac{n}{\log^3 n} } \\
    &\overset{(b)}{\geq} 1 - \round{1 - n^{-c'}}^{\frac{n}{\log^3 n}}\\
    &\overset{(c)}{=} 1 - o(1)
\end{align*}

Here $(a)$ follows from Lemma 2.
$(b)$ holds because Lemma 1 implies that $\prob{F_i \round{\delta} \mid L} \geq \prob{F_i \round{\delta}} - \prob{L^C} \geq n^{-c} - o(1/n)$.
Finally, $(c)$ follows from $c' < 1$.
Similarly,  $\prob{\bigcup_{j \in T_2} \bar{G}_j \round{\varepsilon} \mid L} = 1 - o(1)$ with the same reasoning. Substituting these back into (\ref{eq:independent-breaker}), it follows that $\prob{\bigcup_{i \in C_1, j \in C_2} E_{ij}} = 1 - o(1)$.
\end{proof}


%% file: discussion.tex

Our analysis provides a geometric interpretation for the failure of sym-SDP for asymmetric cases, showing that the hyperplane $w^{\star \top} x = 0$ associated with sym-SDP cannot separate clouds of degree profiles.
Thus, to achieve exact recovery with an SDP, our analysis suggests that the SDP should mimic the MLE. 


Consider the problem in term of the MLE that we should mimic. Let $E$ be the number of edges, $E_1, E_2$ be the number of edges within the first and second communities, respectively, and let $E_{12}$ be the number of edges that connect vertices in different communities. Notice that $E = E_1 + E_2 + E_{12}$.
Additionally, define $H = \binom{n / 2}{2}$ and $R = (n / 2)^2$.

The likelihood of a particular partition is
\begin{align*}
    \mathcal{L}(G) &= p_1^{E_1} (1 - p_1)^{H - E_1} p_2^{E_2} (1 - p_2)^{H - E_2} q^{E_{12}} (1 - q)^{R - E_{12}} \\
    \log \mathcal{L}(G) &= E_1 \log p_1 + (H - E_1) \log (1 - p_1)
    + E_2 \log p_2 + (H - E_2) \log (1 - p_2)
    + E_{12} \log q + (R - E_{12}) \log (1 - q) \\
    &= E_1 \psi_1 + E_2 \psi_2 + C
\end{align*}
where $\psi_1 = \log \round{\frac{p_1}{\round{1 - p_1}} \cdot \frac{\round{1 - q}}{q}}$, $\psi_2 = \log \round{\frac{p_2}{\round{1 - p_2}} \cdot \frac{\round{1 - q}}{q}}$. Note that $\psi_1, \psi_2$ only depend on the problem parameters $p_1, p_2$ and $q$.
Here, $C$ is some constant that does not depend on the partition. Hence, this optimization problem is equivalent to maximizing $E_1 \psi_1 + E_2 \psi_2$ over all valid partitions $\sigma$.

Since our analysis suggests that the objective function of the SDP should match the objective for the MLE, we set $X^{\star} = x^{\star } x^{\star \top}$ to be a rank-$1$ matrix, setting $x^{\star}_i = a \mathbbm{1}\{i \in C_1\} + b \mathbbm{1}\{i \in C_2\}$, where $a = \psi_1$ and $b = -\psi_2$.
This way,
\begin{align*}
    \langle A, X^\star \rangle &= 2 \squarebrack{\psi_1^2 E_1 - \psi_1 \psi_2 E_{12} + \psi_2^2 E_2} \\
    &= 2 \squarebrack{\psi_1^2 E_1 - \psi_1 \psi_2 \round{E - E_1 - E_2} + \psi_2^2 E_2} \\
    &= 2 \squarebrack{\round{\psi_1^2 + \psi_1 \psi_2} E_1 + \round{\psi_2^2 + \psi_1 \psi_2} E_2 - E \psi_1 \psi_2} \\
    &= 2 \round{\psi_1 + \psi_2} \squarebrack{\psi_1 E_1 + \psi_2 E_2} + D
\end{align*}
where $D$ is a constant that does not depend on the partition. Our parameter ordering assumption $p_1 > p_2$ ensures that $\psi_1 + \psi_2 > 0$ so that $\langle A, X^{\star} \rangle$ is equivalent to the likelihood up to scaling and shifting.

We therefore impose constraints on the variable matrix $X$ which are consistent with the desired optimal solution $X^{\star}$, and which encourage the prescribed block structure. These considerations lead to the following SDP:
\begin{equation} \label{eq:our-sdp}
\begin{array}{ll@{}ll}
 \hat{X}_{\mathrm{SDP}} = \arg\max_X  &  \langle A, X \rangle &\\
\text{subject to}& X \succeq 0 \\
& X_{ii} \leq a^2,\; \forall i \in \intset{n} \\
& X_{ij} \geq ab,\; \forall i, j \in \intset{n} \\
& \langle \mathbf{I}, X \rangle = \frac{n}{2} \round{a^2 + b^2} \\
& \langle \mathbf{J}, X \rangle = \frac{n^2}{4} \round{a + b}^2 
\end{array}
\end{equation}


Since our work shows that the failure of sym-SDP can be tied to the geometry of the Genie-aided estimator, it is now natural to view this problem again through the lens of the Genie-aided estimator.
In the case that we observe $d_u = \round{d_1, d_2}$, we predict that $u$ belongs to $C_1$ if
\begin{equation*}
    p_1^{d_{1}} \round{1 - p_1}^{n / 2 - d_{1}}
    \; q^{d_{2}} \round{1 - q}^{n/2 - d_{2}}
    \geq q^{d_{1}} \round{1 - q}^{n / 2 - d_{1}} 
    \; p_2^{d_{2}} \round{1 - p_2}^{n / 2 - d_{2}}.
\end{equation*}
Equivalently, by taking logs, we obtain the condition
\begin{align*}
    &
    d_{1} \log p_1
    + \round{\frac{n}{2} - d_{1}} \log \round{1 - p_1}
    + d_{2} \log q
    + \round{\frac{n}{2} - d_{2}} \log \round{1 - q} \\
    &\geq 
    d_{1} \log q
    + \round{\frac{n}{2} - d_{1}} \log \round{1 - q}
    + d_{2} \log p_2
    + \round{\frac{n}{2} - d_{2}} \log \round{1 - p_2}.
\end{align*}
Rearranging, the Genie-aided estimator classifies $u$ as belonging to $C_1$ if
\begin{align*}
    d_1 \log \round{\frac{p_1}{\round{1 - p_1}} \cdot \frac{\round{1 - q}}{q}}
    + d_2 \log \round{\frac{1 - p_2}{\round{p_2}} \cdot \frac{\round{q}}{1 - q}}
    + \frac{n}{2} \log \round{\frac{1 - p_1}{1 - p_2}}
    &\geq 0 \\
    d_1 \psi_1 - d_2 \psi_2 + \frac{n}{2} \log \round{\frac{1 - p_1}{1 - p_2}}
    &\geq 0,
\end{align*}
and otherwise classifies $u$ as belonging to $C_2$.

It follows that that the correct weight for the separating plane is $\bar{w} = \round{\psi_1, -\psi_2}$ where $\psi_i$ are the same values as those in MLE objective. Also notice that when $p_1 \neq p_2$, there are nonzero term in the condition, so that $\hat{\sigma}\round{u} = \mathrm{sign}\squarebrack{\bar{w}^\top d_u + \frac{n}{2} \log \round{\frac{1 - p_1}{1 - p_2}}}$.

If we consider the failure events $E_{ij}'$ analogously to $E_{ij}$ in (\ref{eq:failure-event}), we have that for $i \in C_1$ and $j \in C_2$,
\begin{align*}
    E_{ij}' &= \set{\round{\psi_2 \sum_{\ell \in C_2} A_{i \ell}
    - \psi_1 \sum_{\ell \in C_1} A_{i \ell}}
    + \round{\psi_1 \sum_{\ell \in C_1} A_{j \ell}
    - \psi_2 \sum_{\ell \in C_2} A_{j \ell}}
    > 0} \\
    &= \set{\bar{w}^\top d_j - \bar{w}^\top d_i > 0}.
\end{align*}

When the parameters are above the IT threshold, with high probability all Genie-aided estimators succeed (see e.g. \cite[Lemma 2.9]{gaudio2024}). Thus,  $\bar{w}^\top d_i + \frac{n}{2} \log \round{\frac{1 - p_1}{1 - p_2}} > 0$ for all $i \in C_1$ and $\bar{w}^\top d_j + \frac{n}{2} \log \round{\frac{1 - p_1}{1 - p_2}} < 0$ for all $j \in C_2$, with high probability. It follows that with high probability
\[\min_{i \in C_1, j \in C_2} \{\overline{w}^{\top} d_j - \overline{w}^{\top} d_i  \} < 0,\]
which implies $\mathbb{P}\left(\bigcup_{i \in C_1, j \in C_2} E_{ij}' \right) = o(1)$.


 Surprisingly, the existing dual certificate proof strategy
 does not readily yield a proof of optimality of the SDP (\ref{eq:our-sdp}). In particular, one natural extension of the form of the certificates in \cite{Hajek2016} is the following lemma:

\begin{lemma} \label{lemma:our-suff-cond}
Suppose there exist $H^\star = \mathrm{diag} \curly{h_i^\star} \geq 0, B^\star \in \mathcal{S}^n , B^\star \geq 0, \eta^\star, \lambda^\star \in \mathbb{R}$ such that
$ S^\star \triangleq H^\star - B^\star - A + \eta^\star I + \lambda^\star J $ satisfies the following:
\begin{itemize}
    \item $S^\star \succeq 0$ with $\lambda_2(S^\star) > 0$
    \item $S^\star x^\star = 0$
    \item $h_i^\star \round{X_{ii}^\star - a^2} = 0$ for all $i$
    \item $B^\star_{ij} \round{X^\star_{ij} - ab} = 0$ for all $i, j$.
\end{itemize}
Then $X^\star$ is the unique solution to the SDP (\ref{eq:our-sdp}).
\end{lemma}

With Lemma \ref{lemma:our-suff-cond}, the only thing left for the optimality of (\ref{eq:our-sdp}) is the certificate that satisfies these conditions with high probability.
Notice that there are many zero-product constraints that limit the structure of the constructed certificate. The constraint $h_i^\star \round{X_{ii}^\star - a^2} = 0$ forces $h_j = 0$ for all $j \in C_2$. Since we need $S^\star x^\star = 0$, it must be the case that for all $i \in C_1$
\begin{align*}
    \round{S^\star x^\star}_i \equiv \round{H^\star x^\star}_i - \round{B^\star x^\star}_i - \round{A x^\star}_i + \round{\eta^\star \mathbf{I} x^\star}_i + \round{\lambda^\star \mathbf{J} x^\star}_i &= 0 \\
    a h_i^\star - \round{B^\star x^\star}_i - \round{A x^\star}_i + \eta^\star a + \lambda^\star \mathbf{1}^\top x^\star &= 0
\end{align*}
Thus, these constraints force the structure of $h_i^\star$ to be the following:
\begin{equation*}
    h_i^\star = \begin{cases}
        \frac{1}{a} \squarebrack{\round{B^\star x^\star}_i + \round{A x^\star}_i - \eta^\star a - \lambda^\star \mathbf{1}^\top x^\star} & i \in C_1 \\
        0 & i \in C_2
    \end{cases}
\end{equation*}

Next, the constraint $B^\star_{ij} \round{X_{ij}^\star - ab} = 0$ forces $B^\star_{ij} = 0$ for all $i, j$ such that $\sigma^\star (i) = \sigma^\star (j)$. Additionally, the requirement $S^\star x^\star = 0$ also forces that for all $j \in C_2$,
\begin{align}
    \round{S^\star x^\star}_j \equiv \round{H^\star x^\star}_j - \round{B^\star x^\star}_j - \round{A x^\star}_j + \round{\eta^\star \mathbf{I} x^\star}_j + \round{\lambda^\star \mathbf{J} x^\star}_j &= 0 \nonumber \\
    - \round{B^\star x^\star}_j - \round{A x^\star}_j + \eta^\star b + \lambda^\star \mathbf{1}^\top x^\star &= 0 \nonumber \\
    \round{B^\star x^\star}_j &= \eta^\star b + \lambda^\star \mathbf{1}^\top x^\star -  \round{A x^\star}_j \nonumber \\
    a \sum_{i \in C_1} B^\star_{ij} &= \eta^\star b + \lambda^\star \mathbf{1}^\top x^\star -  \round{A x^\star}_j \label{eq:column-constraint}
\end{align}
Here, we can view the requirement as a constraint that is imposed on columns $j$. Since the constraint has column structure, we can construct a symmetric matrix $B^\star$ with stripe structure as
\begin{equation*}
    B_{ij}^* = \begin{cases}
    0 & \sigma^\star(i) = \sigma^\star(j) \\
    b_i^* \mathbf{1} \curly{i \in C_2, j \in C_1 } + b_j^* \mathbf{1} \curly{j \in C_2, i \in C_1} & \text{otherwise}
    \end{cases}
\end{equation*}
for some $b_j^\star$ to be determined. With this stripe structure, we can rewrite (\ref{eq:column-constraint}) as
\begin{align*}
    a \cdot \round{\frac{n}{2} \cdot b_j^\star} &= \eta^\star b + \lambda^\star \mathbf{1}^\top x^\star -  \round{A x^\star}_j. \\
    \intertext{Therefore, $b_j^\star$ is determined as}
    b_j^\star &= \frac{2}{a n} \squarebrack{\eta^\star b + \lambda^\star \mathbf{1}^\top x^\star -  \round{A x^\star}_j}.
\end{align*}

It remains to ensure that $S^\star \succeq 0$ with $\lambda_2(S^\star) > 0$.
This is equivalent to ensuring that $\inf_{x \perp \sigma^\star, \lVert x \rVert_2 = 1} x^\top S^\star x > 0$ with high probability.
For any $x$ such that $x \perp \sigma^\star, \lVert x \rVert_2 = 1$,
\begin{align}
    x^\top S^\star x &= x^\top H^\star x - x^\top B^\star x - x^\top A^\star x + x^\top \eta^\star \mathbf{I} x + x^\top \lambda^\star \mathbf{J} x \nonumber \\
    &= x^\top H^\star x - x^\top B^\star x + x^\top \expect{B^\star} x - x^\top A^\star x + x^\top \expect{A} x + \eta^\star + x^\top \lambda^\star \mathbf{J} x - x^\top \expect{B^\star} x - x^\top \expect{A} x \nonumber \\
    &= x^\top H^\star x - x^\top \round{B^\star - \expect{B^\star}} x
    - x^\top \round{A - \expect{A}} x + \eta^\star + \lambda^\star x^\top \mathbf{J} x - x^\top \expect{B^\star} x - x^\top \expect{A} x. \label{eq:requirement-psd}
\end{align}

Since the goal is to make the expression (\ref{eq:requirement-psd}) positive, we can take an approach similar to \cite{Hajek2016} and set $\eta^\star = \lVert A - \expect{A} \rVert + \lVert B^* - \expect{B^\star} \rVert$ to account for the negative contribution from $x^\top \round{B^\star - \expect{B^\star}} x$ and $x^\top \round{A - \expect{A}} x$. Then for any $x$ such that $x \perp \sigma^\star, \lVert x \rVert_2 = 1$,
\begin{align}
x^\top S^\star x &\geq x^\top H^\star x + \lambda^\star x^\top \mathbf{J} x - x^\top \expect{B^\star} x - x^\top \expect{A} x. \label{eq:Sstar}
\end{align}
To summarize, we have the following construction:
\begin{itemize}
    \item $\eta^\star = \lVert A - \expect{A} \rVert + \lVert B^* - \expect{B^\star} \rVert$ 
    \item $h_i^{\star} = \begin{cases}
    \frac{1}{a} \squarebrack{(B^{\star} x^{\star})_i + (A x^{\star})_i - \eta^* a - \lambda^* \mathbf{1}^T x^{\star}} & i \in C_1 \\
    0 & i \in C_2,
    \end{cases}$
    \item $b_j^\star = \frac{2}{a n} \squarebrack{\eta^\star b + \lambda^\star \mathbf{1}^\top x^\star -  \round{A x^\star}_j}$
    \item $B_{ij}^* = \begin{cases}
    0 & \sigma^\star(i) = \sigma^\star(j) \\
    b_i^* \mathbf{1} \curly{i \in C_2, j \in C_1 } + b_j^* \mathbf{1} \curly{j \in C_2, i \in C_1} & \text{otherwise}
    \end{cases}$
\end{itemize}

To show that our construction satisfies the requirement in Lemma \ref{lemma:our-suff-cond}, we want to show that 
\begin{enumerate}
    \item $h_i^\star \geq 0$ for all $i \in C_1$, 
    \item $b_j^\star \geq 0$ for all $j \in C_2$, 
    \item and \eqref{eq:Sstar} holds for all $x$ such that $x \perp \sigma^\star, \lVert x \rVert_2 = 1$.
\end{enumerate}
The first two conditions, along with the fact that $a > b$ imply that for any $i \in C_1$ and $j \in C_2$,
\begin{align}
    \round{B^\star x^\star}_i + \round{A^\star x^\star}_i \geq  \eta^\star a  + \lambda^\star \mathbf{1}^\top x^\star &>  \eta^\star b + \lambda^\star \mathbf{1}^\top x^\star \geq \round{A^\star x^\star}_j \nonumber \\
    \round{B^\star x^\star}_i + \round{A^\star x^\star}_i &>\round{A^\star x^\star}_j \nonumber \\
     b \sum_{\ell \in C_2} b^{\star}_\ell + \squarebrack{\round{A x^\star}_i - \round{A x^\star}_j} &> 0 \label{eq:combined-noneg}
\end{align}

Notice that the first (negative) term $b \sum_{\ell \in C_2} b^{\star}_\ell$ does not depend on $i$ and $j$, since
\begin{align*}
     \sum_{\ell \in C_2} b^\star_\ell &= \frac{2}{a n} \sum_{\ell \in C_2} \squarebrack{\eta^\star b + \lambda^\star \mathbf{1}^\top x^\star -  \round{A x^\star}_\ell} \\
        &= \frac{1}{a} \squarebrack{\eta^\star b + \lambda^\star \mathbf{1}^\top x^\star} - \frac{2}{a n} \sum_{\ell \in C_2} \round{A x^\star}_\ell.
\end{align*}

Notice that the last term is an average of $\frac{n}{2}$ values of $\round{Ax^\star}_\ell$ for $\ell \in C_2$, which concentrates around its expectation, which is equal to $\expect{\round{Ax^\star}_\ell}$ for any representative $\ell \in C_2$.
Observe that for any fixed representative $\ell \in C_2$ and $\tau > 0$ sufficiently small, we have that
\begin{equation} \label{eq:max-mean-relation}
    \max_{j \in C_2} \round{A x^\star}_j \geq \expect{\round{A x^\star}_\ell} + \tau \log n 
\end{equation}
holds with high probability. Let $(p_1, p_2, q)$ lie on the IT threshold, and fix $\tau\round{p_1, p_2, q}$ such that (\ref{eq:max-mean-relation}) holds with high probability for $A$ generated with parameters $p_1, p_2, q$. Then there will be some $\rho = \rho(\tau, p_1, p_2, q) > 0$ such that 
\begin{equation*}
    \max_{j \in C_2} \round{A x^\star}_j \geq \expect{\round{A x^\star}_\ell} + \frac{\tau}{2} \log n 
\end{equation*}
holds with high probability for $A$ generated by feasible, nearby parameters $p_1', p_2', q'$ such that $|p_1' - p_1| + |p_2' - p_2| + |q' - q| < \rho$. Hence, setting $c = \tau(p_1, p_2, q)/2$, we see that in any sufficiently small (feasible) neighborhood of $(p_1, p_2, q)$,
\begin{align}
    \max_{j \in C_2} \round{A x^\star}_j &\geq \expect{\round{A x^\star}_\ell} + c \log n \nonumber \\
    \min_{j \in C_2} \squarebrack{\eta^\star b + \lambda^\star \mathbf{1}^\top x - \round{Ax^\star}_j} &\leq \eta^\star b + \lambda^\star \mathbf{1}^\top x - \expect{\round{A x^\star}_\ell} - c \log n \nonumber \\
    \min_{j \in C_2} b^\star_j = \frac{2}{an}  \min_{j \in C_2} \squarebrack{\eta^\star b + \lambda^\star \mathbf{1}^\top x - \round{Ax^\star}_j} &\leq \frac{2}{an} \squarebrack{\eta^\star b + \lambda^\star \mathbf{1}^\top x - \expect{\round{A x^\star}_\ell} - c \log n} \nonumber \\
    \intertext{where $\ell \in C_2$ is an arbitrary representative.
    By the nonnegativity of $b^\star$,}
    0 \leq \min_{j \in C_2} b^\star_j &\leq \frac{2}{an} \squarebrack{\eta^\star b + \lambda^\star \mathbf{1}^\top x - \expect{\round{A x^\star}_\ell} - c \log n} \label{eq:nonneg-bound}
\end{align}
 Due to the concentration of an average around its expected value, we have that with high probability,
\begin{align*}
    \sum_{\ell \in C_2} b^\star_\ell = \frac{2}{an} \sum_{j \in C_2} \squarebrack{\eta^\star b + \lambda^\star \mathbf{1}^\top x - \round{Ax^\star}_j}
    &\geq \frac{1}{a} \squarebrack{\eta^\star b + \lambda^\star \mathbf{1}^\top x - \expect{\round{A x^\star}_\ell} - \frac{c}{2} \log n} \\
    &= \frac{1}{a} \squarebrack{\eta^\star b + \lambda^\star \mathbf{1}^\top x - \expect{\round{A x^\star}_\ell} - c \log n} + \frac{c}{2a} \log n \\
    &\geq \frac{c}{2a} \log n,
\end{align*}
where the last step follows from (\ref{eq:nonneg-bound}).

Recall that the success of Genie-aided estimators ensures that in the IT-feasible regime, $\squarebrack{\round{A x^\star}_i - \round{A x^\star}_j} \geq 0$ for all $\Theta \round{n^2}$ pairs of $i \in C_1$ and $j \in C_2$. However, those values can be arbitrarily small when approaching the IT threshold. Formally, for any $\varepsilon > 0$, there exist IT-feasible parameters such that $\squarebrack{\round{A x^\star}_i - \round{A x^\star}_j} < \varepsilon \log n$ for some $i \in C_1$ and $j \in C_2$. Therefore, with high probability, there exist $i \in C_1$ and $j \in C_2$ such that
\begin{align*}
    b \sum_{\ell \in C_2} b^\star_\ell + \squarebrack{\round{A x^\star}_i - \round{A x^\star}_j} &\leq b \cdot \frac{c}{2a} \log n + \varepsilon \log n = \round{b \cdot \frac{c}{2a} + \varepsilon} \log n.
\end{align*}

We can ensure $\varepsilon > 0$ to be arbitrarily small by choosing parameters close to the IT threshold, while at the same time $c = \tau(p_1, p_2, q)/2$ is simply a function of the reference point $(p_1, p_2, q)$. Since $b < 0$, taking parameters sufficiently close to the IT threshold
will make the right-hand side negative, which contradicts (\ref{eq:combined-noneg}). Therefore, certificate requirements cannot be satisfied simultaneously.

Notice that the reason of this failure come from the comparison of the magnitude of the sum $b \sum_{j \in C_2} b^\star_j$ against the worst case difference $\round{A x^\star}_i - \round{A x^\star}_j$ over $i \in C_1, j \in C_2$. Meanwhile, our construction of the certificated is mostly forced by the constraints. In particular, there is only a limited choice of $B^\star$ due to the constraint of $S^\star x^\star = 0$ and $B^\star_{ij} \round{X^\star_{ij} - ab} = 0$ while $B^\star$ is required to be symmetric. Hence, this evidence suggests the difficulty of the current dual-certificate proof technique.

While this difficulty is not a proof that the SDP fails to achieve exact recovery, it supports the following conjecture:
\begin{conjecture}
There exist parameters $\alpha_1, \alpha_2, \beta$ such that exact recovery from $G \sim \textsf{SBM}\left(n, \alpha_1, \alpha_2, \beta \right)$ is information-theoretically feasible, yet any SDP fails to achieve exact recovery.
\end{conjecture}